\documentclass[prd,superscriptaddress,
nofootinbib]{revtex4}
\usepackage[colorlinks=true,linkcolor=blue,urlcolor=blue,citecolor=blue]{hyperref}
\usepackage{amsfonts}
\usepackage{amssymb}

\newcommand{\ben}{\begin{displaymath}}
\newcommand{\een}{\end{displaymath}}
\newcommand{\be}{\begin{equation}}
\newcommand{\ee}{\end{equation}}
\newcommand{\bea}{\begin{eqnarray}}
\newcommand{\eea}{\end{eqnarray}}

\begin{document}

\title{A bound  on the nucleon  Druck-term  from chiral  EFT in curved space-time and mechanical stability conditions}
\author{Jambul~Gegelia}
\affiliation{Ruhr-University Bochum, Faculty of Physics and Astronomy,
Institute for Theoretical Physics II, D-44870 Bochum, Germany}
\affiliation{High Energy Physics Institute, Tbilisi State  University,  0186 Tbilisi,
 Georgia}
\author{Maxim V.~Polyakov}
\affiliation{Ruhr-University Bochum, Faculty of Physics and Astronomy,
Institute for Theoretical Physics II, D-44870 Bochum, Germany}
\affiliation{Petersburg Nuclear Physics Institute, 
		Gatchina, 188300, St.~Petersburg, Russia}

\begin{abstract}

Using dispersive representations of the nucleon gravitational form factors, the results for their absorptive parts from chiral effective field theory
in curved space-time, and
the mechanical stability conditions, 
we obtain a model independent inequality for the value of the gravitational $D(t)$ form factor at zero momentum transfer (Druck-term).
In particular, the  obtained inequality leads to a conservative bound on the Druck-term in the chiral limit $D \leq -0.95(9)$. This bound implies the
restriction on the low-energy constant $c_8$ of the effective chiral action for nucleons and pions in the presence of an external gravitational field,
 $c_8\leq -1.1(1)$~GeV$^{-1}$.
For the physical pion mass we obtain a model independent bound  $D\leq -0.20(2)$.

\end{abstract}

\maketitle
\section{Fundamental mechanical properties of the nucleon and  effective chiral Lagrangian in curved space-time}	
\noindent
The fundamental mechanical properties of particles (mass, spin, etc.)
can be related to the linear response functions for the change of the background space-time metric. Thus the basic mechanical properties of particles can be obtained from the
form factors of the energy-momentum tensor (EMT) \cite{Kobzarev:1962wt,Pagels:1966zza}. For the nucleon there are three independent EMT form factors 
\cite{Kobzarev:1962wt,Pagels:1966zza}:\footnote{For definition of the gravitational form factors of hadrons of arbitrary spin see Ref.~\cite{Cotogno:2019vjb}.}
\begin{eqnarray}
\langle p_f, s_f| T_{\mu\nu}(0)| p_i,s_i \rangle &=& \bar u(p_f,s_f) \left[ \bar A(t) P_\mu P_\nu + i \bar J(t) P_{\{\mu} \sigma_{\nu \}\alpha} \Delta^\alpha 
+ \frac 14 \bar D(t) \left(\Delta_\mu \Delta_\nu-\eta_{\mu\nu} \Delta^2\right) \right]  u(p_i,s_i) \, ,
\label{EMTdef}
\end{eqnarray}
where $T_{\mu\nu}(x)$ is the symmetric EMT operator of QCD,
$(p_i,s_i)$ and $(p_f,s_f)$ correspond to the momentum and polarization of the incoming and outgoing nucleons, respectively,  $P=(p_i+p_f)/2$, $\Delta=p_f-p_i$, $t=\Delta^2$, and symmetrisation operation 
is defined as
$X_{\{\mu} Y_{\nu\}}=\frac 12 (X_\mu Y_\nu+X_\nu Y_\mu)$. The values of the  nucleon EMT form factors at zero momentum transfer
\be
{\bar A(0) =\frac{1}{m_N}, \  \ \bar J(0) =\frac{J}{m_N} =\frac{1}{2\, m_N} , \ \ \bar D(0) =\frac{D}{m_N},}
\ee
provide us with three basic mechanical characteristics of the nucleon-- the mass $m_N$, the spin $J=1/2$, and the D-term (or Druck-term\footnote{The name ``$D$-term" is rather technical, it can be traced back to more or less accidental
 notations chosen in Ref.~\cite{Polyakov:1999gs}. Nowadays, given more clear physical meaning of this quantity, we shall call this term as ``{\it Druck-term}" derived from 
German word for pressure.}) $D$. While the mass and spin of the nucleon are well-studied and well-measured quantities, the third mechanical characteristics (the Druck-term)
is more subtle, as it is related to the distribution of the internal  forces inside the nucleon \cite{Polyakov:2002yz} (for a  review see Ref.~\cite{Polyakov:2018zvc}). Important distinguishing
feature of the Druck-term is that to access it one needs variations of the space-time metric such that the resulting Riemann tensor is non-zero. To access the mass and the spin one can
perform the variation of the metric with zero Riemann tensor, e.g. use the variation of the metric corresponding to going to a non-inertial coordinate system.\footnote{Just recall that  the mass of a particle can be measured by studying the bending of the particle's 
track in an external e.m. field, also recall
the classical experiment  with  the Foucault pendulum measuring the Earth's  rotation. Both examples correspond
to a space-time metric induced by going to a non-inertial reference frame.} This qualitative difference explains why the masses and spins are better studied and better 
measured basic properties of particles. 
The discussed difference is clearly seen in the form of the second order effective action for the nucleons (described by the 
Grassmann spinor field  $\Psi(x)$) { and pions (described by $SU(2)$-valued field  $U=u^2$) in a
curved space-time:
\begin{eqnarray}
S_{\rm \pi N} & = & \int d^4x \sqrt{-g}\, \biggl\{
\frac{1}{2} \, \bar\Psi \, i e^\mu_a\gamma^a \nabla_\mu \Psi -\frac{1}{2} \, \nabla_\mu \bar\Psi
\, i e^\mu_a\gamma^a\Psi  -m \bar\Psi\Psi +\frac{g_A}{2}\, \bar\Psi e^\mu_a\gamma^a \gamma_5 u_\mu \Psi  \nonumber\\
&+&  c_1 \langle \chi_+\rangle  \bar\Psi  \Psi  - \frac{c_2}{8 m^2} g^{\mu\alpha} g^{\nu\beta} \langle u_\mu u_\nu\rangle  \left( \bar\Psi \left\{ \nabla_\alpha, \nabla_\beta\right\}  \Psi+
 \left\{ \nabla_\alpha, \nabla_\beta\right\}  \bar\Psi \Psi \right) + \frac{c_3}{2} \, g^{\mu\nu} \langle u_\mu u_\nu\rangle  \bar\Psi  \Psi  \nonumber\\
&+&  \frac{c_8}{8}\, R \bar\Psi \Psi  + \frac{ i c_9}{m} \, R^{\mu\nu} \left( \bar\Psi e_\mu^a \gamma_a \nabla_\nu  \Psi 
- \nabla_\nu  \bar\Psi e_\mu^a \gamma_a  \Psi  \right)   +\cdots 
\biggr\} .
\label{MAction}
\end{eqnarray}
In Eq.~(\ref{MAction}) $g^{\mu\nu}$  and $e^\mu_a$ are the metric (we use  the signature $(+,-,-,-)$) and vielbein gravitational fields \cite{Birrell:1982ix}, respectively,
\begin{eqnarray}
u_\mu & = & i \left[ u^\dagger \partial_\mu u  - u \partial_\mu u^\dagger \right] \,,\nonumber\\
%
\chi_+ & = & 2 B \left[ u^\dagger (s+i p) u^\dagger+u (s-i p) u \right] \,, \nonumber\\
\nabla_\mu \Psi &=& \partial_\mu\Psi +\frac{i}{2} \, \omega^{ab}_\mu \sigma_{ab} \Psi +  \frac{1}{2} \left[ u^\dagger \partial_\mu u  +u \partial_\mu u^\dagger \right]  \Psi, \nonumber\\
\nabla_\mu \bar\Psi &=& \partial_\mu\bar\Psi -\frac{i}{2} \, \bar\Psi \, \sigma_{ab} \, \omega^{ab}_\mu - \frac{1}{2}  \bar\Psi \left[ u^\dagger \partial_\mu u  +u \partial_\mu u^\dagger \right]  ,
\label{CovD}
\end{eqnarray}
where $B$ is related to quark condensate, $s$ and $p$ are external scalar and pseudoscalar source fields, respectively,  $\sigma_{ab}=\frac{i}{2}[\gamma_a, \gamma_b]$, and 
\begin{eqnarray}
\omega_\mu^{ab} &=& -g^{\nu\lambda} e^a_\lambda \left( \partial_\mu e_\nu^b
- e^b_\sigma \Gamma^\sigma_{\mu \nu} \right),\nonumber\\
\Gamma^\lambda_{\alpha \beta} &=& \frac{1}{2}\,g^{\lambda\sigma} \left( \partial_\alpha g_{\beta\sigma}
+ \partial_\beta g_{\alpha\sigma} -  \partial_\sigma g_{\alpha\beta} \right)~, \nonumber\\
R^\rho_{~\sigma\mu\nu} &=& \partial_\mu \Gamma^\rho_{\nu \sigma} -\partial_\nu \Gamma^\rho_{\mu \sigma} + \Gamma^\rho_{\mu \lambda}  \Gamma^\lambda_{\nu \sigma} - \Gamma^\rho_{\nu \lambda}  \Gamma^\lambda_{\mu \sigma}   \,,\nonumber\\ 
R &=& g^{\mu\nu} R^\lambda_{~\mu\lambda\nu}\,.
\label{omega}
\end{eqnarray}
}
%
In the above action the low energy constant $c_8$ is related to the nucleon Druck-term in the chiral limit:
\be
D= c_8 m_N \quad {\rm [chiral\ limit]},
\ee
and the low energy constant $c_9$ gives the leading tree order contribution to the 
slopes of the gravitational form factors $\bar A(t)$ and $\bar J(t)$. From Eq.~(\ref{MAction})
we see that the term responsible for the Druck-term ($\sim c_8$) vanishes in the flat space-time, whereas the terms containing information about
the mass and the spin (the first line in  Eq.~(\ref{MAction})) are non-zero also in flat Minkowski space-time. Clearly, to access the mass and the spin it is
enough to make a metric variation with zero Riemann tensor (just choosing the non-inertial reference frame), whereas the Druck-term can be obtained only by a variation
 with non-trivial curvature. 

The values of the nucleon Druck-term and of the low-energy constant $c_8$ are {\it a priori}  unknown. However, as they are related to distribution of internal forces in the nucleon, these quantities are
restricted by the mechanical stability conditions, see discussions in Refs.~\cite{Polyakov:2018zvc,Perevalova:2016dln,Lorce:2018egm}. In particular, the stability conditions imply
that the nucleon Druck-term (and hence the constant $c_8$) should be negative \cite{Polyakov:2018zvc,Perevalova:2016dln,Lorce:2018egm}. 

The first experimental information on
the nucleon Druck-term \cite{Kumericki:2015lhb,Nature,Kumericki:2019ddg,Dutrieux:2021nlz,Burkert:2021ith,1860995} became available.
In the most recent paper Ref.~\cite{Burkert:2021ith} the Druck-term (quark part) extracted from the DVCS data is negative with
$\sim 9.5\ \sigma$ significance (taking both systematic and statistical uncertainties in quadrature). However, other determinations \cite{Kumericki:2015lhb,Kumericki:2019ddg,Dutrieux:2021nlz}
have obtained much larger systematic uncertainties which preclude to make reliable conclusions about the sign of the Druck-term extracted from DVCS data.

The model and the lattice QCD calculations provide negative 
values of the nucleon Druck-term, see
the review in Ref.~\cite{Polyakov:2018zvc}, and for more recent lattice results see Refs.~\cite{Shanahan:2018nnv,Shanahan:2018pib}. 
In the present paper we strengthen the stability bound $D<0$ using additional information on the chiral expansion 
of the nucleon gravitational form factors obtained in Ref.~\cite{Alharazin:2020yjv}.

\section{Strong force distributions in the nucleon, stability conditions, and dispersion relations}

\noindent
To have  notations coherent with the review \cite{Polyakov:2018zvc} we introduce the gravitational $D(t)$ form factor by rescaling the form factor 
in Eq.~(\ref{EMTdef}) as: $D(t) =m_N \bar  D(t)$. 
The distributions of the pressure $p(r)$ and shear force $s(r)$  can be obtained in terms of $D(t)$ through \cite{Polyakov:2002yz,Polyakov:2018zvc}:
\begin{eqnarray}
\label{Eq:relationSPD}
	s(r)= -\frac{1}{4 m_N}\ r \frac{d}{dr} \frac{1}{r} \frac{d}{dr}
	{\widetilde{D}(r)}, \quad
	p(r)=\frac{1}{6 m_N} \frac{1}{r^2}\frac{d}{dr} r^2\frac{d}{dr}
 	{\widetilde{D}(r)}, \quad
	{\widetilde{D}(r)=}
	\int {\frac{d^3{\bf  \Delta}}{(2\pi)^3}}\ e^{{-i} {\bf  \Delta r}}\ D(-{\bf  \Delta}^2).
\end{eqnarray}
In Refs.~\cite{Polyakov:2018zvc,Perevalova:2016dln,Lorce:2018egm} it was argued that for the stability of a mechanical system the pressure and shear forces should 
satisfy the following inequality:
\be
\label{eq:stab1}
\frac{2}{3} \, s(r) +p(r) \geq 0\,.
\ee
Note that the above stability condition is violated at large $r$ if long-range Coulomb interaction is present (this is not so for the case considered in this paper).
For the discussion of this issue see Refs.~\cite{Varma:2020crx,Metz:2021lqv,PanteleevaMaster}. The stability conditions in the presence of the longe-range 
forces will be discussed in details elsewhere \cite{inpreparation}. {We note that  the Eq.~(\ref{eq:stab1}) is not a consequence 
of the equilibrium equation (EMT conservation), {but rather it is an additional condition for }stability of  the corresponding equilibrium.   }

Using Eq.~(\ref{Eq:relationSPD}) we see that the stability condition of Eq.~(\ref{eq:stab1}) can be written equivalently as:
\be
\label{eq:stab11}
\frac{d}{dr} {\widetilde D(r)} \geq 0,
\ee
implying that the function ${\widetilde D(r)} $ is a monotonically increasing function { which goes to zero for $r\to \infty$.
The monotonically increasing function which { vanishes} at infinity must be negative for $r<\infty$: } 
\be
\label{eq:stab2}
{\widetilde D(r)} \leq 0\,. 
\ee 
For negatively defined function ${\widetilde D(r)} $ we can write the following inequality:
\be
\label{eq:bound}
D \equiv D(0)=\int d^3 {\bf r}\ {\widetilde D(r)} \leq \int_{r\geq R_0} d^3 {\bf r}\ {\widetilde D(r)}, 
\ee 
for an arbitrary distance parameter $R_0$. For  large enough $R_0$, the right hand side of the above equation can  be
computed in the chiral  effective field theory. For that, by applying the dispersion relation for the form factor $D(t)$ without subtraction, one can easily obtain:
\be
\label{eq:stab3}
D\leq \frac 1\pi \int_{4 M_\pi^2}^\infty \frac{dt}{t} \left(1+ R_0 \sqrt{t}\right) e^{-R_0 \sqrt t}\  {\rm Im} D(t+i0).
\ee
At large $R_0$ the integral in this equation is dominated by small $t$, thus it can be computed employing the small $t$ expansion of ${\rm Im} D(t+i0)$. 
The  latter
can be obtained using the chiral effective field theory in curved space-time considered in Ref.~\cite{Alharazin:2020yjv}. To the fourth order
of the chiral expansion the imaginary part of $D(t)$ has the following form:
\begin{eqnarray}
\nonumber
{\rm Im} D(t+i 0)&=& -\frac{3 g_A^2 m_N \left(t-2 M_\pi^2\right)\left(t+4 M_\pi^2\right)}{64 \pi F^2 t^{3/2}}\ {\rm arctg}\left(2 m_N\frac{\sqrt{t-4 M_\pi^2}}{t-2 M_\pi^2}\right) \\
\nonumber
&+&\frac{m_N \sqrt{t-4 M_\pi^2}}{40 \pi F^2 t^{3/2}}
\big[20\ c_1 M_\pi^2  (t+2 M_\pi^2) +c_2 (t^2-3 M_\pi^2 t-4 M_\pi^4)+5\ c_3 (t^2-4 M_\pi^4) \\
\label{eq:imD}
&+&\ \frac{5}{16} \frac{g_A^2}{m_N} (t^2+20 M_\pi^2 t-12 M_\pi^4)
\big].
\end{eqnarray}
This expression is valid for $4 M_\pi^2 \leq t \ll 4 m_N^2$, in its derivation we systematically neglected terms of the order $~\epsilon/(4 m_N^2)$
(here $\epsilon\sim M_\pi^2\sim t$), however the terms of the order
$~\sqrt{\epsilon}/m_N$ were kept. Note that the arctg function is not expanded in $\epsilon/(4 m_N^2)$ to ensure the correct threshold behaviour
of ${\rm Im} D(t+i 0)$ for $t\to 4 M_\pi^2$.

In the above equation $g_A=1.27$ is the nucleon axial coupling constant, $F=0.092$~GeV is the pion decay 
constant,
and the low energy constants (LECs) $c_{1,2,3}$ are the couplings in the standard chiral effective Lagrangian of pions and nucleons of the order two \cite{Gasser:1984yg,Fettes:2000gb}.
The values of these constants $c_{1,2,3}$ can be extracted from the data on $\pi N$ scattering  -  in what follows we use the values obtained in
recent analysis of the $\pi N$ scattering in the framework of the manifestly Lorentz invariant baryon chiral perturbation theory \cite{Yao:2016vbz}, the values of $c_{1,2,3}$ are summarised in Table~\ref{tab:c}.

\begin{table}[t]
\caption{The values of the low energy constants obtained in the analysis of Ref.~\cite{Yao:2016vbz}.
The constants $c_{i}$  are in units of GeV$^{-1}$. The statistical and systematic uncertainties are shown in the first and the second brackets, respectively.}
\begin{center}
\begin{tabular}{|c|c|c|}
\hline
$c_1$ &$ c_2$ & $c_3$\\
\hline
-1.22(2)(2) & 3.58(3)(6)& -6.04(2)(9)\\
\hline
\end{tabular}
\end{center}
\label{tab:c}
\end{table}
The integral in Eq.~(\ref{eq:stab3}),  with the imaginary part of the form factor specified in Eq.~(\ref{eq:imD}), can be computed analytically  if  the chiral limit is considered. The result for the inequality Eq.~(\ref{eq:stab3}) takes the following form:

\be
\label{eq:ineqforc8}
\frac 1 m D^{\rm \small chiral\  lim}= c_8 \leq -\frac{3 g_A^2 }{32 \pi F^2}\frac{1}{R_0}+\frac{3 \left(4 (c_2+5 c_3)+5 g_A^2/m\right)}{80\pi^2 F^2} \frac{1}{R_0^2}+O\left(\frac{1}{R_0^3}\right),
\ee
where $m= 0.8828\pm 0.004$~GeV is the nucleon mass in the chiral limit \cite{Scherer:2012xha}. Now the question is: which value of $R_0$ to choose so that the $O(1/R_0^3)$ corrections can be neglected in Eq.~(\ref{eq:ineqforc8})?
{ {
We follow the standard logic of chiral EFT. The latter provides with a systematic expansion of physical observables in powers of small
parameters $\epsilon\sim M_\pi/\Lambda_\chi, \sqrt{-t}/\Lambda_\chi$, {\it etc.} with $\Lambda_\chi\sim 4\pi F\sim 1$~GeV or $\Lambda_\chi\sim (m_\Delta-m_N)\sim 0.3$~GeV
if contributions of $\Delta$ resonances to the calculated physical observables are expected.
Here we are working with an EFT where the $\Delta$-resonances are not 
included as dynamical degrees of freedom and therefore the expansion of Fourier transformed form factors  in  $1/R_0$ are only suppressed 
 by powers of the nucleon-$\Delta$ mass-difference. By taking $1/R_0\sim 100$~MeV ($R_0\sim 2$~fm and $1/(R_0 (m_\Delta-m_N))\sim 0.3$) 
 we expect that the contributions of higher orders in chiral expansion can be safely dropped within the  accuracy of our calculations.  Considerable improvement of
 the convergency of the chiral expansion (and hence a choice of a smaller value of the parameter $R_0$ in Eq.~(\ref{eq:ineqforc8})) can be achieved 
 by including $\Delta$-s as explicit degrees of freedom in chiral EFT.} 
Additionally we estimated a good choice of the parameter $R_0$ using a model calculation. Following Ref.~\cite{Cebulla:2007ei}, 
we computed the function ${\widetilde D}(r)$ in the Skyrme model in the chiral limit. We observed that  at $r=1.5$~fm 
the deviation from 
 the leading 
large $r$ asymptotics ($\sim 1/r^4$ in the chiral limit) is about 10\% and for $r=2$~fm the deviation is just about 2\%. These model calculations 
give an additional support of our expectation that the chiral expansion in Eq.~(\ref{eq:ineqforc8}) converges rapidly enough for $R_0\sim2$~fm.
In what follows we shall use $R_0=1.5$~fm as the lowest possible value of this parameter in the inequalities (\ref{eq:stab3}), (\ref{eq:ineqforc8}),
and we shall obtain the upper bounds for $c_8$ and $D$ using $R_0=2$~fm.}

Using the values of the low-energy constants from Table~\ref{tab:c} we obtain the upper bounds on $c_8$ shown in Table~\ref{tab:c8} in dependence on
the parameter $R_0$.
\begin{table}[h]
\caption{Lower bounds on the absolute value of LEC $c_8$ obtained with Eq.~(\ref{eq:ineqforc8}). The numbers in brackets are error bars due 
to statistical and systematic uncertainties of LECs $c_{1,2,3}$.}
\begin{center}
\begin{tabular}{|c|c|c| c|c|}
\hline
$R_0$~[fm] & 1.5 & 2 &2.5 &3\\
\hline
$-c_8$~[GeV$^{-1}]\geq $ &$1.64(18)$& $1.10( 10)$&$0.78(6)$ & $ 0.61(4)$\\
\hline
\end{tabular}
\end{center}
\label{tab:c8}
\end{table}
From the table we see that $c_8$ should be {\it negative} and its absolute value is bound from below by a rather sizeable number.
The main uncertainty in our analysis is due to the choice of the parameter $R_0$. To be on the safe side we choose $R_0=2$~fm (see discussion above) for our final estimation 
 $c_8\leq -1.1 (1)$~GeV$^{-1}$.

For non-zero pion mass the integral in Eq.~(\ref{eq:stab3}) can be computed numerically. The resulting bounds on the Druck-term are shown in Table~\ref{tab:D}.
We notice that the obtained bound for the physical pion mass is fulfilled for  all model calculations known to us.
\begin{table}[th]
\caption{Lower bounds on the absolute value of the Druck-term at different values of the pion mass  obtained with Eq.~(\ref{eq:stab3}). The numbers in brackets are error bars due 
to statistical and systematic uncertainties of LECs $c_{1,2,3}$.}
\begin{center}
\begin{tabular}{|c|c|c| c|c|}
\hline
$R_0$~[fm] & 1.5 & 2 &2.5 &3\\
\hline$-D,\ M_\pi=0$ &$\geq1.45(16)$& $\geq0.95(9)$&$\geq0.69(6)$ & $\geq0.54(4)$\\
\hline
$-D,\ M_\pi=M_\pi^{\rm\small phys}$ &$\geq 0.50(6)$& $\geq 0.20(2)$&$\geq 0.081(9)$ & $\geq 0.035(3)$\\
\hline
$-D,\ M_\pi=2 M_\pi^{\rm\small phys}$ &$\geq 0.10(1)$& $\geq 0.021(2)$&$\geq 0.0044(5)$ & $\geq 0.0010(1)$\\
\hline
\end{tabular}
\end{center}
\label{tab:D}
\end{table}

\section{Summary and outlook}
\noindent
An upper bound on the value of the low-energy constant $c_8$ of the chiral effective action in curved space-time is estimated.
 We used dispersive representations by utilizing imaginary parts of the nucleon gravitational form factors, obtained in chiral effective field theory. 
As  a result we obtained 
a model independent inequality for the value of the gravitational $D(t)$ form factor at zero momentum transfer (Druck-term).
 This inequality leads to a conservative bound on the Druck-term in the chiral limit $D \leq -0.95(9)$. The obtained bound implies a
restriction on the low-energy constant $c_8$ of the effective chiral action for nucleons and pions in the presence of an external gravitational field, 
$c_8\leq -1.1(1)$~GeV$^{-1}$.
For the physical pion mass we obtained a model independent bound of $D\leq -0.20(2)$.

The chiral expansion of the imaginary part of the $D(t)$ form factor in Eq.~(\ref{eq:imD}) can have much broader applications than explored here.
For example, it can be used to study in details the distribution of strong forces at nucleon's chiral periphery. Such studies can provide us with
additional insights in the interplay of chiral and confining forces in formation of the nucleon. Corresponding studies will be published 
elsewhere. 

Additionally, the chiral expansion of the absorptive part of the $D(t)$ form factor  (\ref{eq:imD}) paves a way for 
detailed dispersive analysis of the nucleon gravitational form factors based on the Roy-Steiner equations. The corresponding analysis 
for the scalar and electromagnetic form factor were performed in Refs.~\cite{Hoferichter:2012wf,Hoferichter:2016duk}. Using the formalism
developed in Ref.~\cite{Alharazin:2020yjv} and in the current work the corresponding studies can be generalised for the case of gravitational form factors.
Also the calculations presented here will help to improve dispersive evaluation of the nucleon $D(t)$ form factor of Ref.~\cite{Pasquini:2014vua}.

\section*{Acknowledgments}

We are thankful to Julia Panteleeva for interesting discussions of the stability conditions.
This work was supported in part by BMBF (Grant No. 05P18PCFP1),  Georgian Shota Rustaveli National
Science Foundation (Grant No. FR17-354), and  
by the NSFC and the Deutsche Forschungsgemeinschaft through the funds provided to the Sino-German
Collaborative Center TRR110 ``Symmetries and the Emergence of Structure in QCD" (NSFC Grant No. 12070131001, DFG Project-ID 196253076-TRR 110).

\end{document}